\documentclass{emulateapj}
\usepackage{graphicx, amsmath, hyperref, bbold}
\usepackage[lined,boxed]{algorithm2e}

\shorttitle{Polarized Radiative Transfer with MOOGStokes}
\shortauthors{Deen, C.}

\begin{document}

\title{Modification of the MOOG Spectral Synthesis Codes to Account for Zeeman Broadening of Spectral Lines}

\author{Casey P. Deen\altaffilmark{1,2}}

\altaffiltext{1}{Max Planck Institut f\"ur Astronomie, K\"onigstuhl 17, D-69117 Heidelberg, Deutschland}
\altaffiltext{2}{Department of Astronomy University of Texas at Austin, 1 University Station, 78712, Austin, TX, USA}

\begin{abstract}

In an attempt to widen access to the study of magnetic fields in stellar astronomy, I present MOOGStokes, a version of the MOOG one-dimensional LTE radiative transfer code, overhauled to incorporate a Stokes vector treatment of polarized radiation through a magnetic medium. MOOGStokes is a suite of three complementary programs, which together can synthesize the disk-averaged emergent spectrum of a star with a magnetic field.  The first element (a pre-processing script called CounterPoint) calculates for a given magnetic field strength, wavelength shifts and polarizations for the components of Zeeman sensitive lines.  The second element (a MOOG driver called SynStokes derived from the existing MOOG driver Synth) uses the list of Zeeman shifted absorption lines together with the existing machinery of MOOG to synthesize the emergent spectrum at numerous locations across the stellar disk, accounting for stellar and magnetic field geometry.  The third and final element (a post-processing script called DiskoBall) 
calculates the disk-averaged spectrum by weighting the individual emergent spectra by limb darkening and projected area, and applying the effects of Doppler broadening.  All together, the MOOGStokes package allows users to synthesize emergent spectra of stars with magnetic fields in a familiar computational framework.  MOOGStokes produces disk-averaged spectra for all Stokes vectors ($\bf{I},\bf{Q},\bf{U},\bf{V}$), normalized by the continuum.  MOOGStokes agrees well with the predictions of INVERS10 a polarized radiative transfer code with a long history of use in the study of stellar magnetic fields.  In the non-magnetic limit, MOOGStokes also agrees with the predictions of the scalar version of MOOG.
\end{abstract}

\section{Introduction}

In many areas of stellar astrophysics, the effects of magnetic fields are small and may be safely ignored.  However, certain classes of stars (Ap stars, flare stars, active M dwarfs, young stellar objects, etc\ldots) have non-negligible magnetic fields.  Observationally, these strong magnetic fields can significantly affect the equivalent widths of many spectral features normally used to determine stellar physical properties.  A model spectrum with physical parameters $\left( T_{\rm eff}, \log g, \mathrm{[Fe/H]} \right)$ but which omits the effect of the magnetic field will not match the observed spectrum of a magnetic star with the same physical parameters.  More perniciously, a non-magnetic synthetic spectrum with different physical parameters will likely provide a better fit to the observed data, injecting a bias into any study of these parameters or quantities derived from these parameters (spectral type, age, mass etc\ldots).

The increasing sensitivity of current and future infrared spectrographs (CRIRES \citep{Kaeufl2004}, TripleSpec \citep{Wilson2004}, SpeX \citep{Rayner2003} XSHOOTER \citep{Vernet2011}, IGRINS \citep{Yuk2010}, GMTNIRS \citep{Lee2010}, etc\ldots) permit observations of cooler, more embedded objects.  Unfortunately, magnetic fields affect infrared spectra more than visible spectra (the magnitude of the magnetic broadening grows as $\lambda^2$, whereas Doppler broadening scales with $\lambda$).  Magnetic fields affect line shapes in high resolution spectra, and affect equivalent widths of strong lines visisble in low resolution spectra, potentially biasing studies of objects only accessible at infrared wavelengths. Therefore, accurate studies of magnetic stars require a spectral synthesis code which can handle magnetic effects.

\citet{Zeeman1897} qualitatively described the splitting (in wavelength and polarization) of a spectral line under the influence of an external magnetic field. \citet{Hoenl1925} developed the quantum mechanical formulas to describe the Zeeman effect on the spectrum of a parcel of emitting/absorbing material under the influence of a uniform magnetic field as a function of magnetic field strength, geometry, quantum mechanical properties of the transition, and Stokes ($\bf{I}$, $\bf{Q}$, $\bf{U}$, and $\bf{V}$) polarizations.  In a stellar photosphere, the picture is not so simple.  The observed spectrum of a star with a magnetic field is a complicated combination of light produced at different depths of the photosphere, in various polarizations, from different locations across the stellar disk, and at different orientations to the magnetic field.  To complicate matters further, the propagation of polarized light in a magnetic medium requires careful attention to the Stokes parameters.  The evolution through 
the stellar atmosphere of each Stokes parameter depends on the other Stokes parameters as well as on polarized opacities, requiring any code hoping to solve this system of coupled differential equations to treat each of the Stokes parameters as vector quantities.  Radiative transfer codes developed for the synthesis of large spectral regions (MOOG \citep{Sneden_PhD}, SPECTRUM \citep{Gray1994}, Synspec \citep{Hubeny1985}, Synth \citep{Piskunov1992}, etc\ldots), use various algorithms \citep{Feautrier1964, Edmonds1969} to solve the equation of radiative transfer and calculate only the emergent intensity (Stokes $\bf{I}$), and do not account for the interplay between the Stokes vectors.

Motivated by the study of magnetic fields in sunspots, \citet{Unno1956} offered a solution to the Stokes vector equation of radiative transfer through a stellar atmosphere for a normal Zeeman triplet.  \citet{Rachkovsky1962} amended these basic equations to include magneto-optical effects with the addition of the Faraday-Voigt anomalous dispersion profile.  \citet{Beckers1969} then generalized the theory to anomalous Zeeman patterns.  Further investigations into the radiative transfer of polarized radiation through a magnetic medium largely focused on improving computation speed via mathematically complex matrix calculations \citep{Auer1977, Rees1989}, calculating effects due to deviations from local thermodynamic equilibrium (LTE) \citep{DeglInnocenti1976, Auer1977, SocasNavarro2000}, and determining a quantum mechanical basis \citep{DeglInnocenti1972} for the basic equations \citep{Unno1956, Rachkovsky1962}.  Many of these early formulations of the radiative transfer were useful primarily for detailed
studies of single (or small numbers of) absorption lines.  Building on this early work, there are several more recent codes that account for the effects of magnetic fields and polarized radiative transfer (COSSAM \citep{Stift2000}, Zeeman2 \citep{Landstreet1988}, and Synthmag \citep{Piskunov1999}) which have been used to study stars with magnetic fields \citep{Weiss2000, Valenti2001, Stift2001, Landstreet2008, Silvester2012}.

MOOG \citep{Sneden_PhD} is a widely used one-dimensional LTE radiative transfer code with a suite of drivers often used to analyze stellar spectra.  The MOOG driver \texttt{synth} uses stellar atmosphere models together with atomic and molecular line parameters to produce high resolution synthetic emergent spectra.  In this paper, I present a customization of MOOG called MOOGStokes, which permits MOOG to account for the major effects of Zeeman splitting of spectral lines.  MOOGStokes traces the polarized Stokes components though a magnetic stellar photosphere with a uniform magnetic field, producing a disk-averaged spectrum suitable for comparison with observed spectra.  Building the vector radiative transfer package into the familiar framework of MOOG lowers the potential barrier into studies of magnetic fields to an existing broad community of stellar spectroscopists.   Additionally, while Zeeman broadening of absorption lines in infrared spectra can make accurate determinations of $T_{\rm eff}$ and $\log 
g$ impossible for scalar codes (codes which do not solve the Stokes vector equation of radiative transfer), armed with a polarized radiative transfer Zeeman code, the same data can not only constrain temperature and surface gravity more accurately, but also provide a measure of magnetic field strength.  Finally, all spectral synthesis codes make certain assumptions (e.g. their chosen analytic approximation for solving the equation of radiative transfer) and a code which makes different assumptions from other polarized radiative transfer codes, or calculates relevant quantities in different manners can serve as a foil to help elucidate the consequences of those assumptions and the robustness of results \citep{Wade2001}.  In the subsequent paper I give a brief introduction to the theoretical concepts involve in polarized radiative transfer in \S \ref{sec:theory}.  In \S \ref{sec:procedure}, I describe the algorithmic structure of the MOOGStokes suite of software.  \S \ref{sec:verification} describes the 
results of various verification tests and benchmarks, and in \S \ref{sec:discussion}, I discuss possible applications of MOOGStokes, including an illustration of $T_{\rm eff}$ bias caused by neglecting the effects of magnetic fields.

\section{Theoretical Background}
\label{sec:theory}

Before describing the MOOGStokes algorithm in detail, I summarize the relevant theoretical concepts and formulas involved in accounting for the effects of magnetic fields in a synthetic spectrum.  This summary is not intended to be an exhaustive theoretical introduction, but should be sufficient to allow discussion of the algorithms in MOOGStokes.  There are excellent and thorough discussions of this material in \citet{DeglInnocenti1976}, \citet{Rees1989}, and \citet{Piskunov2002} (hereafterward referred to as PK02).

\subsection{The Zeeman Effect}

The Zeeman effect describes the behavior of spectral absorption or emission lines under the influence of an external magnetic field.  In the absence of a magnetic field, the states associated with different magnetic quantum numbers (${\bf M}_J$) of an atomic state (corresponding to the eigenvalues of the angular momentum vector ${\bf J}$ projected along the ${\bf z}$ axis) are degenerate in energy.  An external magnetic field breaks the degeneracy between the eigenstates into $2J+1$ sublevels (denoted by ${\bf M}_J=J, J-1,$\ldots$,-(J-1),-J$).  For a transition between two atomic states, the total number of Zeeman components into which a spectral line will split depends on the ${\bf J}$ of the upper state, ${\bf J}$ of the lower state, and the electric dipole selection rules ($\Delta J = \pm1, 0$, $\Delta J \neq 0$ if $J=0$, $\Delta M_j = \pm 1,0$, $P_f = -P_i$). The shift in energy for a given eigenvalue ${\bf M}_J$ is given by $\Delta E=g\mu_{B}{\bf B}{\bf M_J}$, where $g$ is the Land\'{e} factor, $\mu_{B}
=\frac{e\hbar}{2m_e}$ is the Bohr magneton, and ${\bf B}$ is the strength of the magnetic field.  The shift in energy for a photon emitted between state $E_{\rm up} \left({\bf M}_{J\rm up}\right)$ and $E_{\rm low} \left({\bf M}_{J\rm low}\right)$ is therefore $\Delta E_{\gamma} = \Delta E_{\rm up} - \Delta E_{\rm low} = \mu_{B} {\bf B}\left(g_{\rm up} {\bf M}_{J\rm up} - g_{\rm low} {\bf M}_{J\rm low}\right)$.  Zeeman components with $\Delta {\bf M_J} = 0$ are known as $\pi$ components, and correspond classically to a charge oscillating along the axis of the magnetic field.  $\pi$ components emit radiation linearly polarized the direction of the magnetic field.  Zeeman components with $\Delta {\bf M_J} = 1 (-1)$ are known as $\sigma$ components, and correspond classically to a charge in a right (left) hand circular orbit around the magnetic field vector (according to the IAU definition of Stokes V).  $\sigma$ components emit right (left) hand circularly polarized radiation along the field lines as well as 
linearly polarized radiation in transverse directions.  In the following discussion, quantities related to $\pi$ components are denoted by a $p$ subscript, while quantities related to $\sigma_+, \sigma_-$ components are denoted by $\left(b\right)$lue and $\left(r\right)$ed, signifying the direction of wavelength shift.  Equation \ref{eq:normalizations} (adapted from equations 1-3 in section $4^{16}$ of Condon \& Shortley (1935)) describes the relative intensities $\left(A_{b,p,r}\right)$ of Zeeman components as a function of $\Delta {\bf J}$ and $\Delta {\bf M_J}$ \citep{OrnsteinBurger1924, Hoenl1925} of the initial (lower) state.

\begin{equation} \label{eq:normalizations}
   A_{b,p,r} = 
   \begin{cases}
      M^2 & \Delta J = 0, \pi \\
      \frac{1}{4} \left( J \pm M\right)\left(J\mp M+1\right) & \Delta J = 0, \sigma\\
      \left(J + 1\right)^2 - M^2 & \Delta J = +1, \pi \\
      \frac{1}{4} \left( J \mp M + 1\right)\left(J \mp M + 2\right) & \Delta J = +1, \sigma \\
      J^2 - M^2 & \Delta J = -1, \pi \\
      \frac{1}{4} \left( J \pm M\right)\left(J \pm M - 1\right) & \Delta J = -1, \sigma \\
   \end{cases}
\end{equation}

\subsection{Polarized Radiative Transfer}

While the Zeeman effect describes the effect of a magnetic field on electric dipole transitions and the resultant radiation, still more theory is required to describe the transfer of that radiation through a magnetized stellar photosphere.  In the following discussion, I adopt the geometrical convention described in Figure \ref{fig:angles} (derived from Figure 2 of PK02), where $\gamma$ is the angle the local magnetic field makes with the observer's line of sight, where $\chi$ is the clocking angle of the magnetic field projected onto the plane of the sky as measured from the ${\bf x}$ direction, and where $\theta$ is the viewing angle measured between the line-of-sight and the local normal to the stellar surface.  The different polarizations of the Zeeman component require the use of the Stokes vector ${\bf I}=\left(I,Q,U,V\right)^{\rm T}$, where $I$, $Q$, $U$, and $V$ describe respectively the total intensity, two orthogonal linear polarization intensities, and one circular polarization intensity.  The $^{\
rm T}$ emphasizes that ${\bf I}$ is a column vector.  The equation of radiative transfer now takes vector form:

\begin{figure}
 \begin{center}
    \includegraphics[width=9cm]{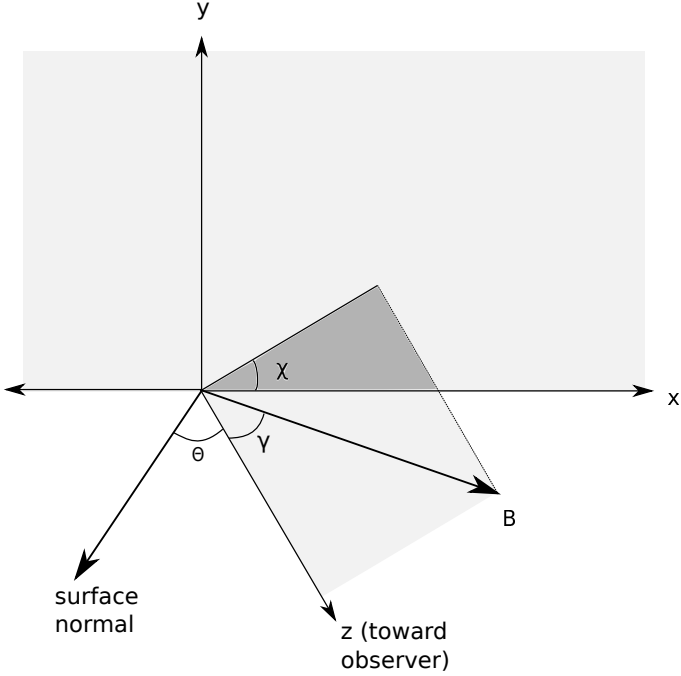}
  \end{center}
  \caption{Definition of angles used in MOOGStokes.  $\gamma$ is defined as the angle between the line-of-sight and the magnetic field $B$, while $\chi$ is the clocking angle of the magnetic field projected onto the plane of the sky, as measured from the ${\bf x}$ direction.  The coordinate system  is usually defined so that $\bf{y}$ points toward north and $\bf{z}$ points toward the observer.  The viewing angle $\theta$ is measured between the line-of-sight and the local surface normal.}
  \label{fig:angles}
\end{figure}

\begin{equation} \label{eq:radtran}
\mu \frac{d{\bf I}}{d\tau}=-{\bf K}\cdot{\bf I} + {\bf J}
\end{equation}

where $\mu=\cos \theta$, $\tau$ is the optical depth, ${\bf K}$ is the opacity matrix, and ${\bf J}$ is the emission vector, each computed at the wavelength of interest.  The opacity matrix ${\bf K}$ (Equation \ref{eq:opacitymatrix}, first introduced by \citet{Unno1956} and later modified by \citet{Rachkovsky1962}) describes the interaction between the intensities of different Stokes components.

\begin{equation} \label{eq:opacitymatrix}
\begin{aligned}
   {\bf K} &= \kappa_C {\bf 1} + \kappa_0 \Phi \\
   {\bf K} &= \left(\begin{array}{cccc}
            \kappa_I & \kappa_0\Phi_Q & \kappa_0\Phi_U & \kappa_0\Phi_V \\
            \kappa_0\Phi_Q & \kappa_I & \kappa_0\Psi_V & \kappa_0\Psi_U \\
            \kappa_0\Phi_U & -\kappa_0\Psi_V & \kappa_I & \kappa_0\Psi_Q \\
            \kappa_0\Phi_V & \kappa_0\Psi_U & -\kappa_0\Psi_Q & \kappa_I \\
            \end{array} \right)
\end{aligned}
\end{equation}

where $\kappa_C$ is the continuum opacity, $\kappa_0$ is the line opacity due to non-hydrogen absorbers/emitters, and $\kappa_I = \kappa_C + \kappa_0\Phi_I$ is the total opacity.  The $\Phi$ matrix is comprised of $\Phi$ (Equation \ref{eq:phi_opacity}) and $\Psi$ (Equation \ref{eq:psi_opacity}) matrix elements:

\begin{equation} \label{eq:phi_opacity}
\begin{aligned}
  \Phi_I &= \frac{1}{2}\left(\phi_p \sin^2\gamma \frac{1}{2}\left(\phi_r+\phi_b\right)\left(1+\cos^2\gamma\right)\right) \\
  \Phi_Q &= \frac{1}{2}\left(\phi_p - \frac{1}{2}\left(\phi_r+\phi_b\right)\right)\sin^2\gamma\cos 2\chi \\
  \Phi_U &= \frac{1}{2}\left(\phi_p - \frac{1}{2}\left(\phi_r+\phi_b\right)\right)\sin^2\gamma\sin 2\chi \\
  \Phi_V &= \frac{1}{2}\left(\phi_r - \phi_b\right)\cos \gamma
\end{aligned}
\end{equation}

\begin{equation} \label{eq:psi_opacity}
\begin{aligned}
  \Psi_Q &= \frac{1}{2}\left(\psi_p - \frac{1}{2}\left(\psi_r+\psi_b\right)\right)\sin^2\gamma\cos 2\chi \\
  \Psi_U &= \frac{1}{2}\left(\psi_p - \frac{1}{2}\left(\psi_r+\psi_b\right)\right)\sin^2\gamma\sin 2\chi \\
  \Psi_V &= \frac{1}{2}\left(\psi_r - \psi_b\right)\cos \gamma
\end{aligned}
\end{equation}

The $\Phi$ matrix elements (Equations \ref{eq:phi_opacity}) are comprised of absorption profiles ($\phi_{b,p,r}$, Equation \ref{eq:phi_bpr}), while the $\Psi$ matrix elements (Equations \ref{eq:psi_opacity}) are composed of anomalous dispersion profiles ($\psi_{b,p,r}$, Equation \ref{eq:psi_bpr}) resulting from magneto-optical effects.  SynStokes calculates $\phi$ and $\psi$ profiles for all Zeeman components that contribute significantly at the current wavelength.  All $\kappa$, $\Phi_{I,Q,U,V}$, $\Psi_{Q,U,V}$, $\phi_{b,p,r}$, and $\psi_{b,p,r}$ implicitly depend upon the current wavelength of interest.

\begin{equation} \label{eq:phi_bpr}
  \phi_{b,p,r}\left(\lambda\right) = \sum_{b,p,r}^{N_{b,p,r}} A_{b,p,r} H\left(a, v-\frac{\Delta\lambda_{b,p,r}}{\Delta\lambda_{\rm Dopp}}\right)
\end{equation}

\begin{equation} \label{eq:psi_bpr}
  \psi_{b,p,r}\left(\lambda\right) = 2\sum_{b,p,r}^{N_{b,p,r}} A_{b,p,r} F\left(a, v-\frac{\Delta\lambda_{b,p,r}}{\Delta\lambda_{\rm Dopp}}\right)
\end{equation}

The $A$ coefficients reflect the relative strengths of different Zeeman components (Equation \ref{eq:normalizations}), and are normalized so that: 

\begin{equation} \label{eq:normalization}
\sum_b^{N_b}A_{b}=\sum_p^{B_p}A_{p}=\sum_r^{N_r}A_{r}=1
\end{equation}

In the nonmagnetic case, there are no wavelength shifts, meaning $\phi_{b,p,r}=\phi_0$ (the absorption profile of the unsplit line), and all $\Phi_{Q,U,V}$ and $\Psi_{Q,U,V}$ are zero due to the opposite signs of the $\pi$ and $\sigma$ components.  In $\Phi_I$, however, the $\pi$ and $\sigma$ components add, giving $\Phi_I = \frac{1}{2}\left(\phi_0 \sin^2 \gamma + \phi_0\left(1+\cos^2\gamma\right) \right) = \phi_0$, and the opacity matrix $\bf{K}$ collapses to the non-magnetic scalar case ($\bf{K} = \kappa_C + \kappa_0 = \kappa_I$)

$H$ and $F$ are respectively the Voigt and Faraday-Voigt functions which describe the absorption and anomalous dispersion profiles as a function of line damping coefficients $\left(a\right)$ and distance from the absorption line center $\left(v\right)$ in terms of the Doppler width for the current absorber species.

The $\Phi$ matrix is also used to calculate the emission vector ${\bf J}$, given by Equation \ref{eq:emissionvector}, assumes both the continuum and line emission are in local thermodynamic equilibrium, and calculates $S_{\rm continuum} = S_{\rm lines} = B\left(\lambda,T_j\right)$, where $B$ is the Planck function, and $T_j$ is the temperature of the $j^{th}$ layer of the atmosphere.

\begin{equation} \label{eq:emissionvector}
{\bf J} = \kappa_C {\bf S}_{\rm continuum} {\bf e}_0 + \kappa_0 {\bf S}_{\rm lines} \Phi {\bf e}_0
\end{equation}
where ${\bf e}_0 = \left(1, 0, 0, 0\right)^T$ is a column vector.

As described in the subesquent section, MOOGStokes now uses the quantities and variables discussed here to solve the Stokes vector equation of radiative transfer (Equation \ref{eq:radtran}) and calculate the emergent spectrum, tracing the Stokes vectors from the base of the photosphere to the top using the Diagonal Lambda Element Operator (DELO) method.

\section{Description of Algorithm}
\label{sec:procedure}

MOOGStokes contains three packages.  A absorption line pre-processor, a MOOG driver, and a disk-integration post-processor.  The pre-processor (named CounterPoint, see Section \ref{sec:CounterPoint}) calculates the number, polarization, and wavelength shift of Zeeman components into which each line splits, given a magnetic field strength.  The MOOG driver (named SynStokes, see Section \ref{sec:MOOGStokes}) takes the Zeeman-split line list together with a model atmosphere and magnetic field geometry, and calculates the emergent spectrum at many different locations across the stellar disk.  The post-processing algorithm (named DiskoBall, see Section \ref{sec:DiskoBall}) then weights the emergent spectrum of each location on the disk by its projected area and limb darkening, applies a Doppler shift due to stellar rotation, and calculates the disk-averaged spectrum as seen by an observer.  Algorithm \ref{alg:moogstokes} shows a pseudocode representation of the suite of three programs.

\subsection{CounterPoint: Absorption Line Pre-Processor}
\label{sec:CounterPoint}
I have developed a Python code called CounterPoint to pre-process absorption lines and account for the Zeeman effect prior to input into MOOG.  Given a magnetic field strength and quantum mechanical constants for the transition (${\bf J}_{\rm up}, {\bf J}_{\rm low}, g_{\rm up}, g_{\rm low}, \log gf$), the program calculates the number and polarization into which the line will split, the relative intensities of the components, and the component energy (wavelength) shifts.  From an initial linelist retrieved from VALD \citep{Kupka2000} CounterPoint produces an entry in a MOOG-readable line list for each Zeeman component, containing the following information: central wavelength, relative oscillator strength, atomic species and ionization state, excitation potential, damping factors if known (i.e. $\Gamma_{\rm vdW},\:\Gamma_{\rm Stark},\: \Gamma_{\rm Rad}$), and change in magnetic quantum number $\left(\Delta {\bf M}_J\right)$.  Because the intensity of a spectral line is directly proportional to the oscillator 
strength, I change the $\log gf$ value of each Zeeman component to match the relative intensities predicted by quantum mechanics \citep{OrnsteinBurger1924, Hoenl1925}, similar to common practices in studies of hyperfine structure.  I perform the normalization (see Equation \ref{eq:normalization}) here so that the sum of the oscillator strengths of all like-polarized components for a line equals the oscillator strength of the unsplit line.  The first part of algorithm \ref{alg:moogstokes} describes the logic of the CounterPoint program.

As an example, CounterPoint calculates the Zeeman splitting of a singly ionized iron line under the influence of a $5.0$ kG magnetic field in the following manner:  The $4923.927 \AA$ Fe line is a dipole transition between a $^6S_{\frac{5}{2}}$ lower state (${\bf J}=\frac{5}{2}$) and a $^6P_{\frac{3}{2}}$ upper state (${\bf J}=\frac{3}{2}$), with an oscillator strength of $\log gf= -1.320$ and a lower state energy of $2.891$ eV.  The Land\'{e} $g$ factors of the lower and upper states are $2.0$ and $2.4$, respectively.  The magnetic field will split the lower state into $6$ levels, and the upper state into $4$ levels.  Application of the electric dipole selection rules determines that there will be four $\pi$ components, and eight $\sigma$ components.  The relative intensities and relative oscillator strengths of the $\pi$ and $\sigma$ components are given in Table \ref{tab:zeeman_example}.

\begin{deluxetable*}{|l|l|l|l|l|l|}
\tablecaption{Zeeman components of the $4923.927 \AA$ Fe II line split by a 5.0kG Magnetic Field.\label{tab:zeeman_example}}
\tablehead{
\colhead{Type} &
\colhead{$\lambda$} &
\colhead{E.P. (eV)\tablenotemark{a}} &
\colhead{$\Delta {\bf M_J}$} &
\colhead{Rel. Int.} &
\colhead{$\log gf$}}
\startdata
$\sigma_{+}$ & 4923.7799 & 2.891\tablenotemark{b} & $\frac{1}{2} \Rightarrow \frac{3}{2}$ & 1 & -2.621\\
$\sigma_{+}$ & 4923.8025 & 2.891\tablenotemark{b} & $-\frac{1}{2} \Rightarrow \frac{1}{2}$ & 3 & -2.144\\
$\sigma_{+}$ & 4923.8251 & 2.891\tablenotemark{b} & $-\frac{3}{2} \Rightarrow -\frac{1}{2}$ &6 & -1.843\\
$\sigma_{+}$ & 4923.8478 & 2.891\tablenotemark{b} & $-\frac{5}{2} \Rightarrow -\frac{3}{2}$ &10& -1.621\\
$\pi$ & 4923.8930 & 2.891\tablenotemark{b} & $\frac{3}{2} \Rightarrow \frac{3}{2}$ & 4 & -2.019 \\
$\pi$ & 4923.9157 & 2.891\tablenotemark{b} & $\frac{1}{2} \Rightarrow \frac{1}{2}$ & 6 & -1.843 \\
$\pi$ & 4923.9383 & 2.891\tablenotemark{b} & $-\frac{1}{2} \Rightarrow -\frac{1}{2}$ & 6 & -1.843 \\
$\pi$ & 4923.9610 & 2.891\tablenotemark{b} & $-\frac{3}{2} \Rightarrow -\frac{3}{2}$ & 4 & -2.019 \\
$\sigma_{-}$ & 4924.0062 & 2.891\tablenotemark{b} & $\frac{5}{2} \Rightarrow \frac{3}{2}$ &10 & -1.621 \\
$\sigma_{-}$ & 4924.0289 & 2.891\tablenotemark{b} & $\frac{3}{2} \Rightarrow \frac{1}{2}$ & 6 & -1.843 \\
$\sigma_{-}$ & 4924.0515& 2.891\tablenotemark{b} & $\frac{1}{2} \Rightarrow -\frac{1}{2}$ & 3 & -2.144 \\
$\sigma_{-}$ & 4924.0741&2.891\tablenotemark{b} & $-\frac{1}{2} \Rightarrow -\frac{3}{2}$ & 1 & -2.621 \\
\enddata
\tablenotetext{a}{Excitation Potential of the lower state}
\tablenotetext{b}{Since the splitting of the lower state energy level will not appreciably affect the populations in each state, CounterPoint does not modify the excitation potential for each component.}
\end{deluxetable*}

\subsection{SynStokes: Polarized Radiative Transfer Calculator}
\label{sec:MOOGStokes}

I describe here the general framework of SynStokes, the FORTRAN driver I added to the MOOG spectral synthesis programs.  The second part of algorithm \ref{alg:moogstokes} describes in pseudocode the algorithm for the MOOG driver `SynStokes':  SynStokes (SynStokes.f\footnotemark[1]) begins by reading in a parameter file (Params.f\footnotemark[2]).  The parameter file describes (among other common MOOG inputs) the number of regions into which the stellar surface will be divided, the desired starting and ending synthesis wavelength, the desired model atmosphere file, and the location of the linelists.  MOOGStokes then reads in the model atmosphere (Inmodel.f\footnotemark[2]) and list of absorption lines (Inlines.f\footnotemark[2]).  After calculating the absorber number density (Eqlib.f\footnotemark[3]), it calculates the populations of absorbers at each layer of the atmosphere, line center opacities (Nearly.f\footnotemark[3]), and polarized opacities (CalcOpacities.f\footnotemark[1]).  SynStokes then creates a 
list of lines which contribute significantly to the opacity (Wavegrid.f\footnotemark[1]), neglecting lines which are too weak due to insufficient opacity in the supplied model atmosphere (e.g. an O II line will not contribute significant opacity in a cool atmosphere, as negligible amounts of oxygen will be ionized).  SynStokes will finely sample the region around strong lines to resolve their shapes, and coarsely sample regions with no strong lines.  Then the program divides the stellar surface into different regions.  For each region, the program calculates the local orientation of the magnetic field relative to the observer (CalcGeom.f\footnotemark[1]), and calculates the emergent spectrum (ComplexVoigt.f\footnotemark[1] Spline.f\footnotemark[1], SplineDriver.f\footnotemark[1], Curfit.f\footnotemark[1], DELOQuad.f\footnotemark[1]), one wavelength point at a time, writing the each spectrum to the output file.  The program then moves on to the next region.  To keep track of the additional variables required 
by polarzied radiative transfer, MOOGStokes adds or modifies the following COMMON blocks to the existing MOOG: Angles.com\footnotemark[1], Atmos.com\footnotemark[2], Linex.com\footnotemark[2], and Stokes.com\footnotemark[1].

\footnotetext[1]{New FORTRAN file unique to MOOGStokes}
\footnotetext[2]{existing FORTRAN file slightly modified to accomodate MOOGStokes}
\footnotetext[3]{existing FORTRAN file used without modification}

In the current implementation, SynStokes adopts a uniform radial geometry for the magnetic field, primarily for its simplicity.  In future versions of the code, the user will be able to specify other magnetic field geometries, as well as multi-temperature atmospheres (as in the case of a stellar spot).

\begin{algorithm}
\SetAlgoLined
\caption{MOOGStokes Algorithm}
\label{alg:moogstokes}
$\bf{CounterPoint}$\;
Read in magnetic field strength $\bf{B}$, $\lambda_{\rm Start, Stop}$\;
Read in VALD line list\;
\For{$i=1 \to n_{\rm lines}$}{
   Compute $n_{\rm Zeeman}$ components for line $i$\;
   Compute $\Delta\lambda$ for each component\;
}
Write Zeeman split lines to Moog-readable line list\;
\BlankLine
$\bf{SynStokes}$\;
Read in parameter file ($n_{\rm regions}$, $\lambda_{\rm Start, Stop}$)\;
Read in model atmosphere (\S \ref{sec:Atmosphere})\;
Read in line list (\S \ref{sec:inlines})\;
Solve Boltzmann/Saha eqns for each species, layer (\S \ref{sec:eqlib})\;
Calculate line center opacities for each transition layer\;
Divide stellar surface into regions (\S \ref{sec:tiles})\;

\For{$i=1 \to n_{\rm regions}$}{
   Compute $\theta_i,\gamma_i,\chi_i$ (\S \ref{sec:tiles})\;
   $\lambda = \lambda_{\rm Start}$\;
   \While{$\lambda \le \lambda_{\rm Stop}$}{
       \For{$j=1 \to n_{\rm layers}$}{
          Compute continuum quantities for layer $j$\;
          Compute line quantities for layer $j$\;
          Compute opacity matrix ${\bf K}$ for layer $j$\;
          Compute emission vector ${\bf J}$ for layer $j$\;
       }
       Compute optical depths $\tau_l$, $\tau_C$ (\S \ref{sec:OpticalDepths})\;
       Compute emergent ${\bf I}, {\bf Q}, {\bf U}, {\bf V}$ w/ DELO alg (\S \ref{sec:DELO_int})\;
       $\lambda += \Delta\lambda$\;
   }
   Save region spectrum\;
}
\BlankLine
$\bf{DiskoBall}$\;
Read in spectra, $n_{\rm regions}$, $\theta$, $\gamma$, $\chi$\;
\eIf{regions are annular}{
   Compute composite spectrum using \emph{rtint} algorithm \citep{Valenti1996}
}{
   \For{$i=1 \to n_{\rm regions}$}{
      Compute limb darkening, proj. area coefficients and doppler shift for region $i$\;
      Add spectrum $i$ to composite spectrum scaling by coefficients\;
 }
}
Save composite spectrum\;
\end{algorithm}

\subsubsection{Model Atmosphere}
\label{sec:Atmosphere}

The model atmosphere gives SynStokes the temperature, pressure, and density profiles of the photospheric region as functions of the optical depth.  I opt not to include the effects of magnetic pressure on the atmospheric structure.  Future versions of the code may explicitly address the issue of magnetic pressure.

\subsubsection{Absorption Line List}
\label{sec:inlines}
The original scalar version of MOOG, using a routine in Inlines.f, reads information regarding absorption lines from a file, where each line of the file contains the following information: Wavelength of the transition $\lambda$ (in $\AA$ or $\mu$m), atomic (or molecular) species, energy of the lower state (in electron volts), oscillator strength of the transition, van der Waals damping coefficient, and molecular dissociation energy (only for molecular lines).  I modified the Inlines routine to accept three additional paramters: $\Delta {\bf M}_J$, $\Gamma_{\rm Rad}$, and $\Gamma_{\rm Stark}$.  The change in angular momentum $\Delta {\bf M}_J$ as calculated by CounterPoint, allows SynStokes to identify $\pi$ and $\sigma_-, \sigma_+$ Zeeman components.  $\Gamma_{\rm Rad}$ and $\Gamma_{\rm Stark}$ are damping constants related to radiative and Stark broadening, respectively.

\subsubsection{Equilibrium Calculations and Line Center Opacities}
\label{sec:eqlib}
SynStokes then uses the existing machinery of MOOG to solve the classical Boltzmann and Saha equations to calculate the number of absorbers of each species and lower energy state at each layer of the atmosphere, assuming LTE.  The number of absorbers in turn allows the calculation of the line-center opacity $\kappa_0$ for each line in the line list.  A non-LTE calculation would include radiative and density effects \citep{SocasNavarro2000}, and would slightly affect the line opacity and source function, but the LTE approximation is frequently made by other stellar radiative transfer codes \citep{Landstreet1988, Stift2000, Piskunov2002}.

\subsubsection{Dividing the Stellar Surface into Different Regions}
\label{sec:tiles}
The emergent spectrum of a magnetic star is a function of the geometry of the magnetic field, the stellar photosphere, and the observing angle.  SynStokes provides the user two strategies for synthesizing disk-averaged spectra.  The first strategy (described in PK02) divides the stellar surface into a number of approximately equal-area tiles.  The user can specify the inclination and clocking angle of the stellar rotation axis, and can control the tile size by specifying in the MOOG parameter file the number of total tiles and number of latitude belts.  For each tile, SynStokes calculates the angles $\gamma$ and $\chi$ for the center of the tile, as well as the viewing angle $\theta$ (see Figure \ref{fig:angles}).  $\gamma$ and $\chi$ are calculated using the orientation of the local magnetic field (assumed to be uniform and radial, and hence $\theta = \gamma$) relative to the observer.  If the viewing angle $\theta$ implies that the center of the tile is visible to the observer ($\cos\theta > 0$), SynStokes 
calculates the emergent spectrum.  This first strategy produces emergent spectra for all four Stokes parameters, but due to the large number of tiles necessary for calculation of an accurate average flux, can be quite slow.  The second strategy available to the user is to divide the stellar disk into a number of annuli, and calculate the emergent Stokes I and V spectra for each annulus at the stellar equator.  While significantly faster, this strategy only produces disk-averaged spectra for Stokes I and V (due to the azimuthal dependence of Stokes Q and U).

\subsubsection{Calculation of Line Opacities, Opacity Matrix ${\bf K}$, and Source Function ${\bf J}$}
\label{sec:Opacities}

In order to calculate ${\bf K}$ for a wavelength $\lambda$ and atmospheric layer $j$, SynStokes first calculates the total $\phi_{b,p,r}$ (Equations \ref{eq:phi_opacity}) and $\psi_{b,p,r}$ opacities (Equations \ref{eq:psi_opacity}) summed over all lines in the line list which contribute significant opacity at $\lambda$.  As the original scalar version of MOOG contained only a formula for the Voigt profile \citep{Kurucz1970}, I include in SynStokes the algorithm from \citet{Humlicek1982} to calculate both the Voigt and Faraday-Voigt profiles. Equipped with the individual polarized opacities and the geometry of the magnetic field, SynStokes calculates the individual elements of the opacity matrix ${\bf K}$ (see Equation \ref{eq:opacitymatrix}) and of the emission vector ${\bf J}$ (see Equation \ref{eq:emissionvector}).  SynStokes then constructs a sequence of opacity matrices and emission vectors calculated at each layer of the atmosphere.

\subsubsection{Calculation of Optical Depths}
\label{sec:OpticalDepths}

SynStokes must also keep track of the line and continuum optical depths to calculate an emergent spectrum.  The optical depth given in the model atmosphere file is a reference optical depth (often measured at either a reference wavelength, or a Rosseland mean opacity). MOOG converts this reference optical depth to an optical depth at the current wavelength $\lambda$ and atmospheric layer $j$ using $\kappa_C$ and $\kappa_{\rm ref}$.  However, MOOG does not calculate the physical depth through the atmosphere (a quantity required by the DELO algorithm in \S \ref{sec:DELO_int}), so SynStokes must caclculate this quantity.  Adopting $d\tau_l = - \kappa_I dz$ for the line optical depth and $d\tau_C = - \kappa_C dz$ for the continuum optical depth at the current wavelength $\lambda$, SynStokes converts the reference optical depth given in the model atmosphere to a physical depth into the photosphere by integrating the equivalent equation for the reference optical depth $z = -\int_0^{\tau_{\rm ref}}\frac{1}{\kappa_{\
rm ref}}\,\mathrm{d}\tau$.  Calculation of $\kappa_I$ and $\kappa_C$ then allows SynStokes to calculate line and continuum optical depths at the current atmospheric layer and wavelength $\lambda$.

\subsubsection{DELO Integration}
\label{sec:DELO_int}

The scalar version of MOOG employes a formal integrative methodology to obtain contribution functions which are subsequently used to calculate the emergent intensity at a given wavelength \citep{Edmonds1969, Sneden_PhD}.  To account for the exchange of light between different Stokes vectors, it becomes necessary to solve the Stokes vector equation of radiative transfer (Equation \ref{eq:radtran}) through the atmosphere of the star.  The most straight-forward method of solving this system of differential equations is a brute-force Runge-Kutta algorithm \citep{DeglInnocenti1976}.  However, while accurate, Runge-Kutta algorithms are quite computationally intensive.  Even given the numerous folding-length times of Moore's Law between the computers available to \citet{DeglInnocenti1976} and the computers of today, the time required to synthesize all but the smallest of wavelength intervals becomes prohibitively long.  For this reason, I adopt the Diagonal Element Lambda Operator (DELO) method of \citet{
ReesMurphy1987}.  This method slightly modifies equation \ref{eq:radtran}, allowing the propagation of the Stokes ${\bf I}$ vector to be treated as a linear relation between adjacent points in the stellar atmosphere (see Equations \ref{eq:DELO}).  To improve the accuracy of this algorithm, I use the quadratic formula $\left(\mathcal{Z}\right)$ for the emission vector $\bf{S'}$ employed by \citet{Olson1987}, \citet{Kunasz1988}, \citet{SocasNavarro2000}, and PK02.  For a more detailed discussion of the DELO algorithm and definitions of the constants contained in equations \ref{eq:DELO}, I refer the reader to the excellent treatments in \citet{ReesMurphy1987}, \citet{SocasNavarro2000}, and in PK02.  SynStokes sub-samples each decade of $\tau_{\rm ref}$ by steps of $0.05$ dex, to trace each Stokes vector from the base ($\log\tau_{\rm ref} \sim 2$) to the top ($\log\tau_{\rm ref} \sim -5$).  I assume the radiation originating at the base of the photosphere to be initially unpolarized and in LTE (${\bf I}=B\left(T_
j\right), {\bf Q}={\bf U}={\bf V}=0$).  I then calculate the emergent continuum by setting all non-continuum sources of opacity to zero and performing the same DELO integration.  The MOOGStokes implementation of the DELO algorithm uses the ATLAS and LAPACK linear algebra packages \citep{Anderson1999, Blackford2002, Whaley2004}, as well as the Numerical Recipes implementation of cubic splines \citep{Press1992}.

\begin{equation} \label{eq:DELO}
\begin{aligned}
  \mathcal{X}_i \cdot \bf{I}\left(\tau_i\right) &= \mathcal{Y}_i \cdot \bf{I}\left(\tau_{i+1}\right) + \mathcal{Z}_i \\
  \mathcal{X}_i &= \mathbb{1} + \left( \alpha_i - \beta_i\right) \bf{K'}_i \\
  \mathcal{Y}_i &= \left(\epsilon_i\mathbb{1} - \beta_i\bf{K'}_{i+1}\right) \\
  \mathcal{Z}_i &= \gamma_i\bf{S'}_{i-1} + \eta_i\bf{S'}_i + \zeta_i\bf{S'}_{i+1} \\
  \bf{K'} &= \frac{\bf{K}}{\kappa_I} - \mathbb{1} \\
  \bf{S'} &= \frac{\bf{J}}{\kappa_I} \\
\end{aligned}
\end{equation}

\subsubsection{Storing Emergent Spectra}
SynStokes calculates an emergent spectrum for each tile of stellar surface visible to the observer.  SynStokes saves each individual spectrum (in Stokes I, Q, U, V, and continuum), along with a description of its geometry ($\theta, \gamma, \chi$) for post-processing (as described in the subsequent section).

\subsection{DiskoBall: Disk Integration Post-Processor}
\label{sec:DiskoBall}

Once SynStokes has calculated emergent spectra from the stellar tiles visible to the observer, the individual spectra must then be averaged together to produce the final disk-averaged spectrum.  For this purpose, I have created a Python post-processing script called DiskoBall.  DiskoBall reads emergent spectra from the output files created by SynStokes, and combines them into a single disk-averaged spectrum.  If the spectra correspond to tiles covering the entire stellar surface, DiskoBall calculates the disk-averaged spectrum by weighting the flux of each tile by the limb darkening (\S \ref{sec:LimbDarkening}) and projected area of the tile (\S \ref{sec:ProjArea}), and shifting the wavelength by the appropriate Doppler velocity (\S \ref{sec:DoppShift}), given the source geometry (inclination and clocking angle).  DiskoBall then saves a composite spectrum for each Stokes parameter.  If instead, the spectra correspond to annuli, DiskoBall produces a composite spectrum by convolving the spectrum of each 
annulus with a rotation kernel and weighting by annular area and limb darkening (\ref{sec:LimbDarkening})\citep{Valenti1996}.

\subsubsection{Limb Darkening}
\label{sec:LimbDarkening}
I use the simple power-law limb darkening prescription from \citet{Hestroffer1998}, primarily for its simplicity.  As the limb darkening effect is wavelength dependent, Diskoball calculates the limb darkening coefficient for the mean wavelength, and applies it to the entire spectrum.  While newer limb darkening laws can produce more accurate results, the \citet{Hestroffer1998} law provides sufficient accuracy for the initial release of the program.

\subsubsection{Projected Area}
\label{sec:ProjArea}
DiskoBall weights the emergent spectrum coming from each one of the surface tiles by the projected area, as seen by the observer.  The projected area is the surface area ($\mathrm{d}\alpha \cdot \mathrm{d}\phi$) multiplied by the cosine of the viewing angle ($\mu=\cos \theta$).

\subsubsection{Doppler Broadening}
\label{sec:DoppShift}
DiskoBall allows the user to provide a rotational velocity $v$ for the requested composite spectrum.  Together with the source geometry, DiskoBall calculates the appropriate $v \sin i$ for the spectrum from each element on the stellar disk and applies the corresponding red or blue shift, before coadding with the other spectra.  This feature allows the user to re-process a single output of SynStokes with any number of arbitrary stellar rotational velocities.

\section{Verification}
\label{sec:verification}

Before trusting the predictions of any spectral synthesis code, the code must reproduce to high accuracy the calculations of other well-tested spectral synthesis codes under identical input conditions.  To verify that the code can accurately synthesize spectra when no magnetic fields are present, I check the emergent Stokes I calculated by MOOGStokes against that of its predecessor, MOOG.  After successfully proving MOOGStokes introduces no major deviations, I then check the code against the magnetic profiles provided in \citet{Wade2001}.

\subsection{Non-Magnetic Verification}
\begin{figure}
 \begin{center}
    \includegraphics[width=9cm]{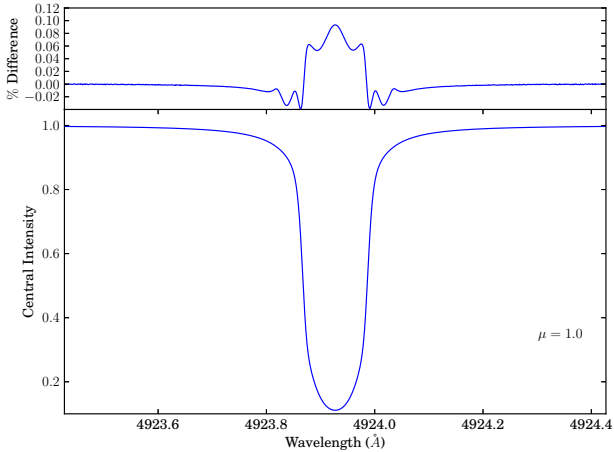}
  \end{center}
  \caption{Comparison of emergent intensities calculated at infinite resolving power by MOOG and MOOGStokes.  The bottom of the figure shows the MOOGStokes emergent spectrum calculated at disk center ($\mu=1.0$) by the $4293.297 \AA$ Fe II line in an Atlas9 model atmosphere ($T_{\rm eff} = 7500K\: \log g = 4.0\: v_{\rm mt}=0\:km\,s^{-1}$) with the magnetic field set to zero.  The top portion shows the difference between the profile produced by MOOGStokes and that produced by MOOG, in units of the continuum.}
  \label{fig:diff_mu_1}
\end{figure}

MOOGStokes uses the existing software framework of MOOG to support a completely different spectral synthesis engine.  To verify that the chassis of MOOG is properly connected to its new engine, I have tested that MOOGStokes produces the same result as the original MOOG in the non-magnetic limit.  For this test, I synthesize emergent central intensity (Figure \ref{fig:diff_mu_1}, $\mu=1.0$) produced by the Fe II line described in Table \ref{tab:zeeman_example} with an ATLAS9 model atmosphere \citep{Castelli2003}.  There are small differences on the order of $0.1\%$ of the continuum.  These differences are likely due to small numerical differences between the DELO and contribution function algorithms, both approximate analytical solutions to the radiation transport equation.  The profiles produced by the two codes agree to a level where errors in the spectrum will be dominated by uncertainties in the parameters of the transitions being modeled.

\subsection{Comparison to \citet{Wade2001}}

\begin{figure}
 \begin{center}
    \includegraphics[width=9cm]{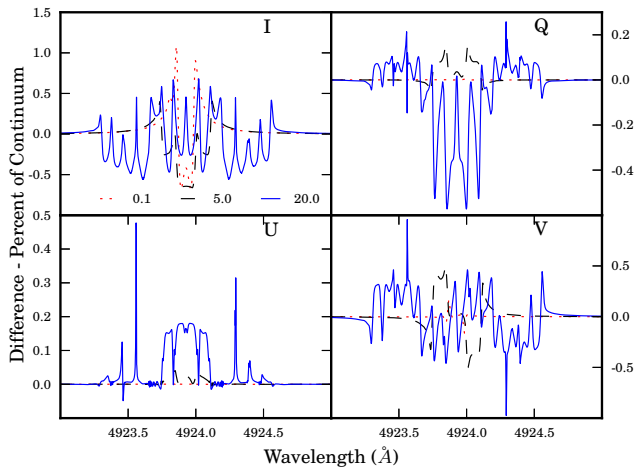}
  \end{center}
  \caption{Comparison between the Stokes profiles of the Fe II line (described in Table \ref{tab:zeeman_example}) produced under identical input conditions (Atlas9 model atmosphere: $T_{\rm eff} = 7500K\: \log g = 4.0\: v_{\rm mt}=0\:km\,s^{-1}$) by MOOGStokes and INVERS10.   Each of the four panels shows the percentage difference (in units of the continuum) between the predictions of MOOGStokes and the profiles shown in Figure 4 of \citet{Wade2001}.  The different lines correspond to different magnetic field strengths (red dotted--$0.1kG$, black dashed--$5.0kG$, blue solid--$20.0kG$).}
  \label{fig:wade_comp}
\end{figure}

The most important test of a polarized radiative transfer code is a synthesis of a transition under the influence of a strong magnetic field.  Figure 4 of \citet{Wade2001} shows Stokes IQUV profiles produced by the Fe II line described in Table \ref{tab:zeeman_example} under the influence a magnetic fields of strength $0.1$, $5.0$, and $20.0kG$, as calculated by the INVERS10 polarized radiative transfer code.  INVERS10 makes use of the more accurate Feautrier algorithm \citep{Feautrier1964, Auer1977}, while MOOGStokes uses the quadratic DELO algorithm \citep{SocasNavarro2000}.  Figure \ref{fig:wade_comp} shows a comparison of the predictions of MOOGStokes with the profiles provided in \citet{Wade2001}.  While the differences between MOOGStokes and INVERS10 ($\sim 0.25\%$) are larger than those between MOOG and MOOGStokes ($\sim 0.1\%$), they are still small enough that other uncertainties in the creation of synthetic spectra (model atmospheres, oscillator strengths, etc\ldots) will dominate the errors in any 
comparisons to real observations.

\section{Discussion}
\label{sec:discussion}

I have described the necessary steps to modify an existing scalar spectral synthesis code to account for the major effects produced by Zeeman splitting and polarized radiative transfer. The result of these modifications, MOOGStokes, is sufficient for the study of the behavior of absorption line shapes and equivalent widths under the influence of changes in the physical parameters of the photosphere ($T_{\rm eff}$, $\log g$, and ${\bf B}$).  I have attempted to make the interface of MOOGStokes similar and complementary to that of the original MOOG program, allowing observers and stellar spectroscopists already familiar with MOOG to make the transition to studying magnetic fields.

As intimated in the introduction, neglecting the effects on spectra of strong magnetic fields can affect conclusions drawn from the spectra.  The damping wings and widths of certain absorption lines are frequently used to determine certain physical parameters (surface gravity, microturbulence, $v \sin i$, etc\ldots).  The equivalent widths of other lines are often used to constrain other parameters (effective temperature, metallicity, etc\ldots).  Weak optically thin lines, which are in the linear portion of the curve of growth change their shapes under the influence of a magnetic field, but do not change appreciably in equivalent width.  Strong optically thick lines change shape as well, but also increase in equivalent width, due to the saturation of the individual Zeeman components in the logarithmic portion of the curve of growth.  While changes in line shape only become noticeable at high spectral resolution, changes in equivalent widths of strong lines affect spectra of all resolutions (and hence the 
aforementioned properties derived from them).

\begin{figure}
 \begin{center}
    \includegraphics[width=9cm]{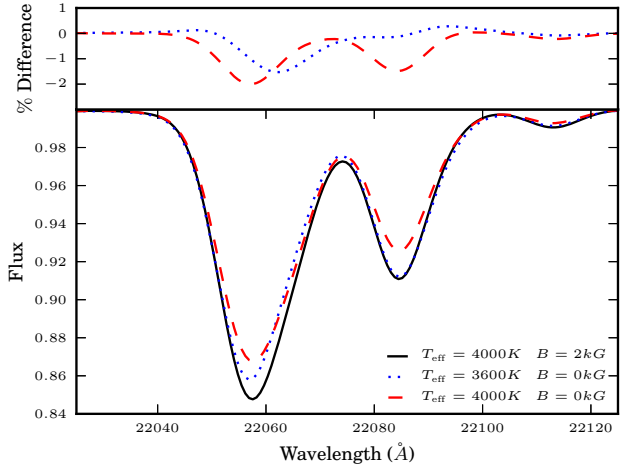}
  \end{center}
  \caption{Comparison between one magnetic ($2kG$) and two non-magnetic spectra. The spectra are shown in the bottom of the figure, and have been convolved down to spectral typing resolution ($R\sim2000$).  The top portion of the figure shows the differences between the magnetic spectrum and the two non-magnetic ones in percentage of continuum.  The spectrum with the same effective temperature (blue dashed line) provides a worse fit to the magnetic spectrum }
  \label{fig:beffect}
\end{figure}

As an illustration of the magnitude of the effect that strong magnetic fields can have on the appearance of the emergent spectrum of a star as observed by a normal spectrograph (Stokes I), Figure \ref{fig:beffect} shows a comparison between three synthetic spectra of the sodium doublet at $2.2\mu$m convolved to $R=\frac{\Delta \lambda}{\lambda} = 2000$.  For late-type stars, the equivalent width of the sodium doublet is often used in determining spectral type and veiling (excess continuum emission due to hot circumstellar dust).  The first spectrum (black solid line) is a spectrum generated with parameters appropriate for a low mass young stellar object ($T_{\rm eff} = 4000K$, $\log g = 4.0$, and average magnetic field strength of $2.0kG$).  Converting effective temperature to spectral type, this corresponds to a spectral type of roughly K7 \citep{MamajekTeff, Luhman2003}.  The second spectrum (red dashed line) shows a spectrum generated with parameters appropriate for a young non-magnetic K7 star ($T_{\rm 
eff} = 4000K$, $\log g = 4.0$, and no magnetic field).  The third spectrum (blue dotted line) is a spectrum of a young non-magnetic M1.5 star ($T_{\rm eff}=3600K$, $\log g = 4.0$, no magnetic field).  The equivalent width of the sodium doublet in the M1.5 star matches the
equivalent width of the magnetic young star better than the K7 star, even though its effective temperature is $400K$ cooler.  Astronomers often determine ages and masses of young stellar objects by comparing their locations on the HR diagram to evolutionary models \citep{Baraffe1998, PallaStahler1999}.  An error in effective temperature of the magnitude displayed in this example can result in errors in the derived age of several million years and errors several tenths of $M_{\Sun}$ in the derived stellar mass, introducing biases into studies of young stellar object properties (i.e. lifetimes of circumstellar disks, initial mass functions, etc\ldots).

While magnetic fields can make determining stellar parameters from individual or small numbers of absorption lines difficult, not all lines are affected equally by the Zeeman effect.  In subsequent investigations, I will use MOOGStokes, along with this fact, as tools to determine physical parameters ($T_{\rm eff},\log g,$ magnetic field strength) of stars with magnetic fields.

\subsection{Further Work and Acknowledgments}

Further versions of the MOOGStokes code will address non-radial, non-uniform magnetic fields, and temperature variations caused by spotting across the disk of the star.

During the development of this code, I became indebted to many experts in radiative transfer, polarized or otherwise.  Chris Sneden, Rob Robinson, Dan Jaffe, John Lacy, Christopher Johns-Krull, Cornelis Dullemond, Oleg Kochukhov, and Juan Manuel Borrero all provided invaluable advice and suggestions.  I wish to also thank the anonymous referee, whose comments and suggestions improved the manuscript.  This work was begun under a NASA USRA SOFIA Grant.  All portions of the MOOGStokes package (CounterPoint, SynStokes, and DiskoBall) are available upon request from the author or from the author's \href{http://www.mpia.de/~deen/}{website}.

\end{document}